\documentclass[conference]{IEEEtran}
\IEEEoverridecommandlockouts    

\usepackage{multirow}
\usepackage{lipsum}
\usepackage{amsmath}
\usepackage{amssymb}
\usepackage[ruled,vlined]{algorithm2e}
\usepackage{float}
\usepackage{graphicx}
\graphicspath{{./Figures/}}
\usepackage[hidelinks]{hyperref}
\usepackage{xcolor}
\usepackage{mdframed}
\definecolor{lightyellow}{RGB}{255,255,200}
\usepackage{cite}
\usepackage{amsfonts}
\usepackage{algorithmic}
\usepackage{textcomp}
\usepackage{soul}
\usepackage{comment}
\sethlcolor{yellow}

\title{\LARGE \bf ARCAS: An Augmented Reality Collision Avoidance System with SLAM-Based Tracking for Enhancing VRU Safety}

 \author{
 	\parbox{\textwidth}{%
 		\centering
 		Ahmad Yehia$^{1}$\textsuperscript{*}, Jiseop Byeon$^{1}$\textsuperscript{*}, Tianyi Wang$^{1}$\textsuperscript{*}, Huihai Wang$^{2}$, Yiming Xu$^{2}$, Junfeng Jiao$^{2}$, Christian Claudel$^{1\dag}$%
 	}%
 	\thanks{\textsuperscript{\dag}Corresponding author: Christian Claudel.}%
    \thanks{\textsuperscript{*}These authors contributed equally to this work.}%
    \thanks{$^{1}$Department of Civil, Architectural, and Environmental Engineering, The University of Texas at Austin, Austin, TX 78712, USA.
 		{\tt\small ahmad.yehia@utexas.edu, jsbyeon@utexas.edu, bonny.wang@utexas.edu, christian.claudel@utexas.edu}}%
 	\thanks{$^{2}$School of Architecture, The University of Texas at Austin, Austin, TX 78712, USA.
 		{\tt\small hw9998@utexas.edu, yiming.xu@utexas.edu, jjiao@austin.utexas.edu}}%
 }

\begin{document}
	
	\maketitle
	\thispagestyle{empty}
	\pagestyle{empty}
	
	\begin{abstract}
Vulnerable road users (VRUs) face high collision risks in mixed traffic, yet most existing safety systems prioritize driver or vehicle assistance over direct VRU support. This paper presents ARCAS, a real-time augmented reality (AR) collision avoidance system that provides personalized spatial alerts to VRUs via wearable AR headsets. By fusing roadside 360° 3D LiDAR with SLAM-based headset tracking and an automatic 3D calibration procedure, ARCAS accurately overlays world-locked 3D bounding boxes and directional arrows onto approaching hazards in the user’s passthrough view. The system also enables multi-headset coordination through shared world anchoring. Evaluated in real-world pedestrian interactions with e-scooters and vehicles (180 trials), ARCAS nearly doubles pedestrians’ time to collision and increases counterparts’ reaction margins by up to 4× compared to unaided eye conditions. Results validate the feasibility and effectiveness of LiDAR-driven AR guidance and highlight the potential of wearable AR as a promising next generation safety tool for urban mobility.
	\end{abstract}
	
\section{Introduction}
Traffic crashes involving vulnerable road users (VRUs) remain a major global safety concern. According to the World Health Organization, pedestrians and cyclists account for more than half of the 1.35 million annual traffic fatalities worldwide \cite{who2019}.  In the United States, pedestrian and cyclist deaths have increased at an alarming rate. During the first half of 2023, 3,373 pedestrians lost their lives on US streets, representing a 19\% increase compared to 2019 \cite{ghsa2024}. Similarly, 1,149 cyclists were killed on US roads in 2023, marking a 4\% increase from 2022. As autonomous vehicles (AVs) become more common in urban areas, the absence of a human driver necessitates alternative methods to interpret vehicle intentions to VRUs \cite{wang2025hlcg, li2026characteristics}. Consequently, there is a growing need for systems that enhance communication and situational awareness for VRUs in mixed traffic environments \cite{ackermans2020, carmona2021}.

To mitigate these challenges, numerous safety technologies have been introduced, including automatic emergency braking systems, smart crosswalks, and raised pedestrian infrastructure~\cite{guo2022study}. However, these technologies still rely on driver compliance, assuming that drivers will act logically and follow traffic laws. Prior studies have demonstrated that external human–machine interfaces (eHMIs), including LED strips, LED screens, and road projections, can compensate the lack of driver cues and support pedestrians in making more effective road-crossing decisions \cite{bazilinskyy2021}. Despite their potential, current eHMI solutions have certain limitations, including difficulties in addressing individual pedestrians \cite{colley2020}, challenges in visibility due to occlusions \cite{dey2022}, and the absence of standardization \cite{tabone2021a}, which can lead to inconsistent interface designs that confuse pedestrians and pose potential safety risks.

A promising recent solution to these challenges involves VRU-vehicle interaction through wearable augmented reality (AR) devices. AR overlays virtual information onto the real world, allowing users to remain aware of their surroundings while receiving safety relevant alerts \cite{tong2021, tabone2021a}. Unlike vehicle-mounted displays, the AR-based approach effectively enables personalized communications to individual VRUs, given that the information is no longer limited to the vehicle exterior \cite{tabone2021b, tran2022}. While AR has great potential for VRU safety, such applications may seem unconventional or impractical to navigate traffic safely \cite{tabone2021b}. The growing adoption of AR technologies, such as Google AR navigation on smartphones, suggests that visual assistance through AR glasses could be a feasible safety solution for VRUs in the near future.

\begin{figure}[!t]
    \centering
    \vspace{6pt}
    \includegraphics[width=\linewidth]{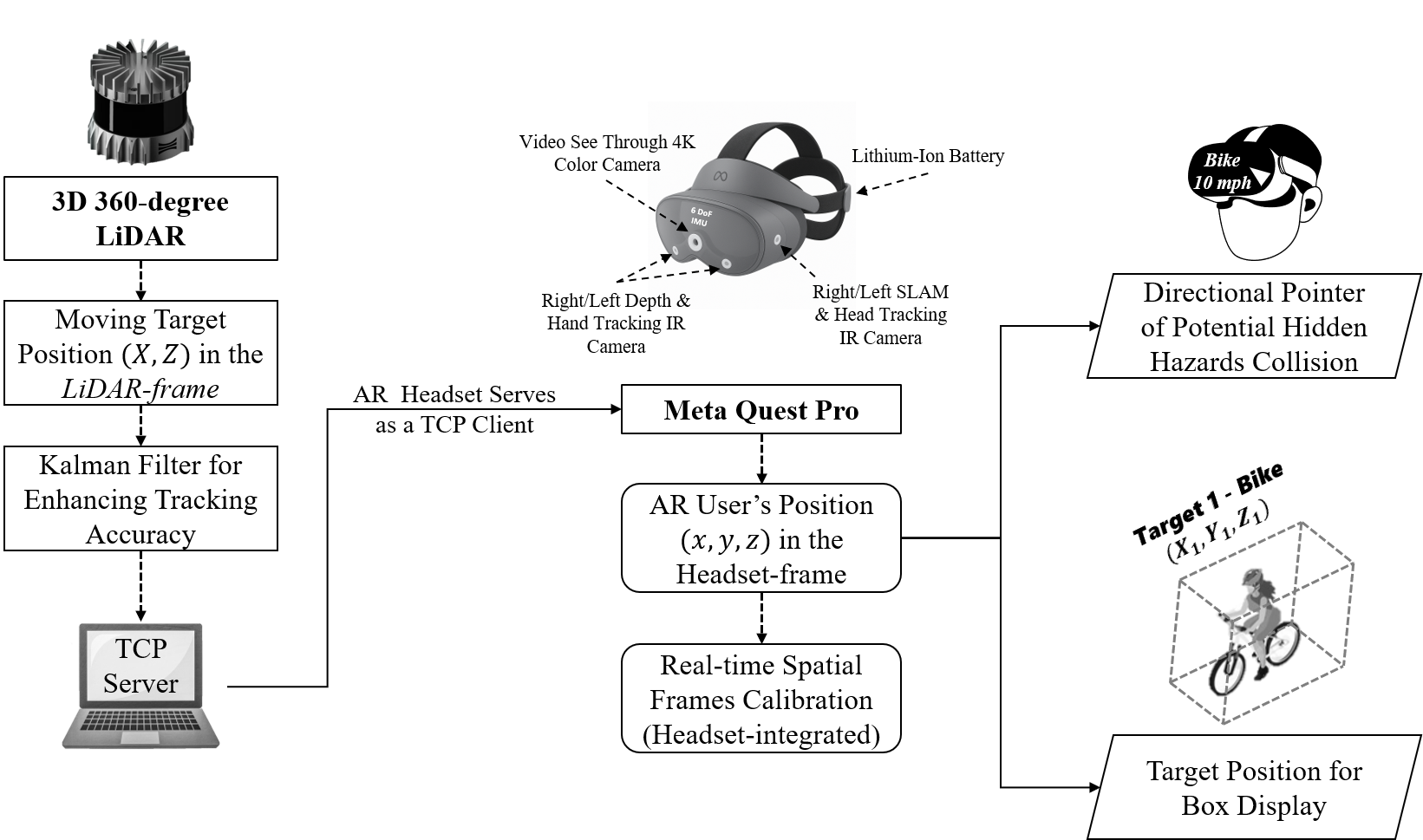}
    \caption{System overview of the proposed ARCAS. The left portion of the figure illustrates the AR headset (a Meta Quest Pro operating in passthrough mode), while the right portion shows the positioning module, consisting of a 3D 360° LiDAR sensor and the associated TCP server.}
    \label{fig:system_framework}
\end{figure}

Most existing AR-based studies have focused on the driver perspective, exploring head-up and windshield displays to highlight nearby pedestrians or cyclists \cite{wiegand2019}. Although a growing number of studies have examined AR interfaces designed for VRUs \cite{nex2014}, many remain conceptual or simulated. Existing solutions are designed to provide road crossing information, such as zebra crossing and directional arrows \cite{hesenius2018, prattico2021}, and to convey collision warning signs to VRUs, including vehicle overlays \cite{tong2021, tran2022}. Only a limited number of studies have explored AR applications in real traffic environments, and these have primarily focused on evaluating interface concepts for conveying information rather than developing fully traffic collision warning system \cite{tabone2021b}. To the best of our knowledge, no prior work has demonstrated a real-time AR collision avoidance system for VRUs in real traffic or examined multi-user AR alignment for coordinated awareness.

To address these research gaps, the primary objective of this paper is to develop an \textbf{A}ugmented \textbf{R}eality \textbf{C}ollision \textbf{A}voidance \textbf{S}ystem (ARCAS), a real-time framework designed to enhance VRU safety across dynamic traffic conditions (Figure \ref{fig:system_framework}). ARCAS integrates roadside 360° 3D LiDAR sensing with wearable AR displays to continuously track surrounding traffic, detect potential collision risks, and deliver immediate spatial alerts. A key strength of the system is its automatic 3D calibration between the LiDAR frame and the AR headset frame, enabling accurate alignment of virtual warnings with real-world objects. Additionally, ARCAS supports multi-headset operation, allowing multiple VRUs or first responders, such as firefighters, to share a unified world coordinate system for coordinated situational awareness in complex environments. More specifically, the main contributions of this novel AR-based approach are summarized as follows:

\begin{enumerate}
    \item \textbf{Wearable AR Visualization and Calibration}: A Meta Quest Pro (MQPro) headset performs real-time functions, including the user's 6 DoF pose using visual-inertial simultaneous localization and mapping (SLAM) \cite{durrantwhyte2006slam}, gaze tracking, and video streaming, while also running the calibration procedure that aligns the LiDAR coordinate frame with the headset reference frame. The headset functions as a TCP client to receive 3D target information from the server and renders spatial cues, such as bounding boxes and directional arrows, to alert users of potential collisions in real time.

    \item \textbf{Multi-User Capability}: ARCAS introduces a multi-headset feature that establishes a shared world coordinate frame, enabling multiple users to maintain coordinated situational awareness in collaborative or emergency scenarios.
    
    \item \textbf{Real-World Evaluation}: Unlike prior AR–VRU concepts evaluated primarily in simulation or controlled environments \cite{tong2021, tabone2021a, tabone2021b}, ARCAS is implemented and tested in real-world traffic conditions, demonstrating its feasibility in dynamic VRU–vehicle interactions.
\end{enumerate}

\section{Related Work}

Prior research has identified human error as the leading cause of road accidents \cite{hendricks2001}. Technologies such as advanced driver assistance systems, adaptive cruise control, and fully AVs aim to improve driver and passenger safety \cite{wang2025impact}; however, VRUs remain at high risk. For example, in interactions with AVs, the implicit communication conveyed solely through vehicle motion may be insufficient to ensure safe VRU–vehicle interactions \cite{prattico2021}. As a result, recent studies have focused on external communication methods and interface systems that explicitly convey a vehicle’s status and intentions to surrounding road users.

\subsection{Vehicle-mounted Interfaces}
Initial pedestrian–vehicle interface designs relied on visual signals mounted on the vehicle exterior. One early concept, “Eyes on a Car” \cite{chang2017}, used animated eyes embedded in the headlights to mimic driver–pedestrian eye contact. The eyes tracked pedestrians to indicate yielding intent or looked forward to signal non yielding behavior. While this design helped pedestrians make faster crossing decisions in virtual experiments, some users reported discomfort due to the Uncanny Valley effect and questioned the system’s reliability \cite{lagstrom2016}.
Similarly, common approach used LED strip-based interfaces installed over the windshield or along the vehicle side. These interfaces used dynamic lighting patterns to signal different vehicle states such as stopping, yielding, or resuming movement \cite{habibovic2018}. Pedestrians generally found these cues intuitive, with side-mounted strips preferred for their clarity. However, some participants questioned the necessity of continuous visual feedback in all situations.
The “Smiling Car” interface \cite{locken2019} was another LED-based design that incorporated anthropomorphic features, transforming a horizontal yellow line into a smile when a pedestrian is detected to indicate yielding intent. Studies involving single pedestrian-vehicle interactions found the interface intuitive and effective in conveying clear messages \cite{deb2018}. However, these vehicle-mounted interfaces remain limited by their reliance on visual signals, which can be difficult to perceive in poor weather, low light conditions, or from a distance.

\subsection{Road Projection-based Interfaces}

To address the limitations of vehicle-mounted displays, researchers have investigated projection-based interfaces that project visual signals directly onto the road surface. One prototype \cite{burns2019} projected parallel lines in front of the vehicle, with line spacing that varies based on vehicle speed. Although this design helped pedestrians make quicker crossing decisions, their attention often shifted to the projection rather than the vehicle itself. Another approach \cite{nguyen2019} used a color changing projection system with three colors, red during normal driving, yellow when slowing for VRUs, and green when it is safe to cross. While these projections improve visibility and build on familiar cues such as crosswalk markings, their effectiveness can degrade under certain road conditions and may increase VRU cognitive load.

\subsection{Smart Road Interfaces}
Beyond vehicle-mounted and projection-based systems, smart road interfaces embed visual signals, such as LED crosswalks, directly into the pavement to convey crossing information to VRUs. These systems are generally perceived as highly reliable and easy for VRUs to interpret; however, the high cost of infrastructure upgrades remains a significant limitation \cite{locken2019}.

\subsection{Wearable Augmented Reality Concepts}
Wearable AR offers many of the capabilities of smartphones while enabling more immersive interaction. Unlike traditional devices, AR overlays virtual content onto the real world, enhancing information display and situational awareness. As consumer grade AR glasses become more accessible, VRUs may soon adopt wearable AR in everyday settings, creating new opportunities for advanced VRU-vehicle interaction paradigms \cite{tabone2021a}.
Wearable AR can function as an intelligent assistant that interprets the surrounding environment, monitors user states (e.g., location, orientation, and gaze), and adapts information to situational needs. Expert interviews in recent work \cite{tabone2021a} highlighted the potential of wearable AR to address scalability challenges in AV–VRU interaction. Accordingly, several studies have investigated AR-based eHMIs. Hesenius et al. \cite{hesenius2018} developed three concepts for pedestrians: one to display a visual walking path, another to highlight safe crossing zones, and a third to convey vehicle intentions and predicted stopping points. Tong and Jia \cite{tong2019} proposed an AR-based system that alerts pedestrians about approaching vehicles to improve situational awareness during crossings.
More recently, Tabone et al. \cite{tabone2021b} introduced a conceptual AR prototype that communicates vehicle intentions and crossing information directly to VRUs through wearable displays. In addition, Lin et al. \cite{lin2024} introduced an AR-based system for real-time pedestrian conflict alerts, but the system was not evaluated in real-world urban environments. While these studies offer promising concepts, most remain limited to theoretical frameworks, AR-interface designs, or single vehicle scenarios with limited empirical validation in complex real-world environments involving multiple traffic hazards. To address these gaps, the present research aims to design an empirical, real-time AR-based collision avoidance system designed to enhance VRU safety in dynamic traffic conditions.

\section{System Design}

\subsection{General Overview}

The ARCAS system is structured around real-time communication between a roadside 3D 360° LiDAR sensor and a wearable MQPro headset, coordinated through a centralized TCP server running on a laptop (specifications detailed in Section IV-A). The ARCAS TCP architecture consists of a TCP client deployed on the AR headset and a corresponding TCP server interfaced with the 3D LiDAR sensor. The system components are illustrated in Figure \ref{fig:system_framework}. Object tracking is managed by the LiDAR sensor, while all computations related to spatial mapping, calibration, and visual-cue rendering are performed locally on the AR headset. The headset software is developed in C\# using the Unity engine, while the server software is implemented in C++. The software workflow operates as follows:

\begin{enumerate}
    \item The client initializes and establishes the headset's local frame.
    \item The client connects to the TCP server and receives initial 3D positional data in the headset's world frame.
    \item A calibration process is performed on the TCP client by capturing user motion simultaneously in both the headset and world frames to compute the coordinate transformation between them.
    \item After calibration, the headset continuously renders target objects (excluding the headset user) within its local frame and displays directional visual cues (e.g., arrows) for targets outside the field of view that may pose a collision risk.
\end{enumerate}

\subsection{Real-Time Coordinate Transformation for AR Visualization}
In this application, all users are assumed to move within a 3D space defined by coordinates $(x, y, z)$. The AR headset internal tracking system, based on SLAM, provides the spatial information of the headset relative to the world frame, denoted as $(X, Y, Z)$. With the introduction of a 3D LiDAR sensor, the transformation between these reference frames is modeled as a full 3D
rotation followed by a translation, as described in Equation (\ref{eq:rotationmatrix}).

\begin{equation} \label{eq:rotationmatrix}
\left[ \begin{array}{c} X \\ Y \\ Z \end{array} \right] =
\left[ \begin{array}{c} X_0 \\ Y_0 \\ Z_0 \end{array} \right] +
\mathbf{R}
\left[ \begin{array}{c} x \\ y \\ z \end{array} \right].
\end{equation}
After the calibration phase, the system computes the optimal values for
$(X_0, Y_0, Z_0)$ and the 3D rotation matrix $\mathbf{R}$. The positioning system, based on a 3D LiDAR sensor, provides real-time measurements of the positions of both headset user and other moving targets. The system applies the transformation defined in Equation (\ref{eq:rotationmatrix}) to each target to accurately render 3D bounding boxes in the headset's world frame.

\subsection{Kalman Filter}
To reduce the measurement noise associated with the external 3D positioning system, a Kalman Filter is employed to estimate both position and velocity within the LiDAR reference frame. In the 3D configuration, the Kalman Filter is extended to estimate full three-dimensional position and velocity, resulting in a six-dimensional state vector and a 6×6 state transition model. In contrast, the AR internal tracker introduces negligible noise and requires no additional filtering.

In this study, the state vector includes position and velocity:
\[
x_k = \begin{bmatrix} 
x(k) \\ y(k) \\ z(k) \\
v_x(k) \\ v_y(k) \\ v_z(k)
\end{bmatrix},
\quad
z_k = \begin{bmatrix}
x_{lidar}(k) \\ y_{lidar}(k) \\ z_{lidar}(k)
\end{bmatrix}.
\]

The discrete time motion equation used for prediction is:
\begin{equation} \label{e:kf1}
x_{k+1} = A x_k + w_k,
\end{equation}
where \( A \) is the state transition matrix, and \( w_k \) is the process noise vector. 

The corresponding observation equation is defined as:
\begin{equation} \label{e:kf2}
z_k = H x_k + v_k,
\end{equation}
where \( H \) is the observation matrix, and \( v_k \) is the observation noise vector:
\[
H =
\begin{bmatrix}
1 & 0 & 0 & 0 & 0 & 0 \\
0 & 1 & 0 & 0 & 0 & 0 \\
0 & 0 & 1 & 0 & 0 & 0
\end{bmatrix}.
\]

The Kalman Filter is applied using the standard prediction and update steps:

\textit{Prediction step:}
\begin{align}
\hat{x}_{k|k-1} &= A \hat{x}_{k-1|k-1}, \\
P_{k|k-1} &= A P_{k-1|k-1} A^\top + Q.
\end{align}

\textit{Update step:}
\begin{align}
K_k &= P_{k|k-1} H^\top \left( H P_{k|k-1} H^\top + R \right)^{-1}, \\
\hat{x}_{k|k} &= \hat{x}_{k|k-1} + K_k \left( z_k - H \hat{x}_{k|k-1} \right), \\
P_{k|k} &= \left( I - K_k H \right) P_{k|k-1}.
\end{align}
where $\hat{x}_{k|k-1}$ and $\hat{x}_{k|k}$ denote the predicted and updated state estimates, $P_{k|k-1}$ and $P_{k|k}$ are the corresponding covariance matrices, and $K_k$ is the Kalman gain. $Q$ and $R$ denote the process and observation noise covariances, respectively.

The noise covariance matrices are defined as follow:
\[
Q = \mathrm{diag}(\lambda^2, \lambda^2, \lambda^2, \mu^2, \mu^2, \mu^2), \quad
R = \mathrm{diag}(\sigma^2, \sigma^2, \sigma^2).
\]
where \( \lambda \) and \( \mu \) represent uncertainties in position and velocity dynamics, respectively, and \( \sigma \) is determined by the accuracy of the positioning sensor used in the experiment.

As a result, these filtered position and velocity estimates \( \hat{x}_{k|k} \) are then used to update the AR visualization in real time, ensuring smoother and more reliable motion tracking despite the observation noise.

\subsection{Automatic Headset Calibration}
The system performs automatic calibration by aligning coordinates generated by the 3D LiDAR with those tracked by the headset internal system. This process relies on synchronized 3D position data from both the AR headset and the LiDAR, captured as the user moves within the LiDAR detection range. Using this paired data, the calibration task is formulated as a 3D optimization problem to estimate the rotation matrix $\mathbf{R}$ and translation offset $(X_0, Y_0, Z_0)$, as defined in Equation (\ref{e:calib}). 
\begin{equation} \label{e:calib}
\min_{\mathbf{R},\,X_0,\,Y_0,\,Z_0}
\sum_{i=1}^{n}
\left\|
\begin{bmatrix} X(i) \\ Y(i) \\ Z(i) \end{bmatrix}
-
\mathbf{R}
\begin{bmatrix} x(i) \\ y(i) \\ z(i) \end{bmatrix}
-
\begin{bmatrix} X_0 \\ Y_0 \\ Z_0 \end{bmatrix}
\right\|^{2}.
\end{equation}

To estimate the rotation matrix $\mathbf{R}$, the centroids of each dataset are subtracted to center the coordinates. This process transforms the calibration problem into a special case of the Orthogonal Procrustes problem \cite{schonemann1966generalized}, which seeks the optimal orthogonal transformation to align two point sets:
\begin{equation}
\min_{\Omega \;\; \text{s.t.} \;\; \Omega^T \Omega = I} 
\left\| \Omega A - B \right\|_F,
\end{equation}
where $A$ and $B$ are centered coordinate matrices, $\Omega$ is the sought 3D rotation matrix, and $\|\cdot\|_F$ is the Frobenius norm.

In practice, an initial estimate of the linear transformation between the two coordinate sets is obtained, after which a singular value decomposition is applied to enforce orthogonality. If the unconstrained transformation is written as $T = U \Sigma V^T$, the closest proper 3D rotation matrix in the Frobenius norm sense is given by:
\begin{equation}
    \mathbf{R} = U V^T.
\end{equation}

This procedure ensures that $\mathbf{R}$ is a valid rotation matrix in $\mathrm{SO}(3)$, providing a consistent alignment between the LiDAR and headset coordinate frames.

\subsection{Off Field-Of-View Target Pointer}
In real-world scenarios, vehicles or VRUs posing a collision risk may fall outside the headset user field of view (FoV). To address this limitation, ARCAS incorporates a directional pointer cue to alert the user to off-screen targets approaching or in potential collision path. These  cues help the AR headset users identify threats and respond accordingly, even when visual occlusions or limited FoV prevent direct line of sight.
To determine whether a target lies outside the FoV (106°), the system computes the angle between the headset user forward-facing direction vector and the vector pointing from the ego user to the moving target. If this angle exceeds a predefined FoV threshold, a directional pointer is activated and displayed at the screen edge to indicate the target's direction. Otherwise, no pointer is rendered, assuming the user can see the target directly. This approach is inspired by video games, where off-screen objects are represented as markers pointing toward their relative direction in a top-down or first-person view.

\subsection{Algorithms}
The server, running on a machine connected to the 3D LiDAR, handles object tracking, and data transmission. The client, running on the AR headset, manages system initialization, coordinate calibration, and real-time visualization. The pseudocode for each component is presented below.

\begin{algorithm}[!ht]
\caption{Server Pseudocode Logic} \label{alg:alg1}
\begin{algorithmic}
\STATE \textsc{3D LiDAR Initialization}
\STATE \hspace{0.5cm}Initialize 3D LiDAR driver at 20 Hz and start streaming point clouds

\STATE \textsc{Main Loop}
\STATE \hspace{0.5cm}Capture current 3D LiDAR point cloud
\STATE \hspace{0.5cm}Segment dynamic objects and detect up to $n$ targets
\STATE \hspace{0.5cm}\textbf{For each detected target $T_i$:}
\STATE \hspace{1cm}Apply 3D Kalman Filter to estimate position $\mathbf{p}_i = (x_i, y_i, z_i)$
\STATE \hspace{1cm}Estimate velocity $\mathbf{v}_i = (v_{xi}, v_{yi}, v_{zi})$

\STATE \textsc{TCP Communication}
\STATE \hspace{0.5cm}If client connected, send $\mathbf{p}_i$ and $\mathbf{v}_i$
\STATE \hspace{0.5cm}Else, apply timeout/reconnect logic to handle temporary network dropouts
\end{algorithmic}
\end{algorithm}

\begin{algorithm}[!ht]
\vspace{6pt}
\caption{Client Pseudocode Logic (AR Headset)} \label{alg:alg2}
\begin{algorithmic}
\STATE \textsc{Initialization}
\STATE \hspace{0.5cm}Connect to TCP server

\STATE \textsc{Calibration Phase}
\STATE \hspace{0.5cm}Get ego coordinates via SLAM $\mathbf{p}_{\text{ego}} = (x, y, z)$
\STATE \hspace{0.5cm}Get corresponding LiDAR coordinates $\mathbf{P}_1 = (X_1, Y_1, Z_1)$ from server
\STATE \hspace{0.5cm}\textit{// First detected target corresponds to ego user}
\STATE \hspace{0.5cm}Estimate 3D rotation matrix $\mathbf{R}$ and translation offset $\mathbf{t} = (X_0, Y_0, Z_0)$

\STATE \textsc{Main Loop}
\STATE \hspace{0.5cm}Get updated ego coordinates via SLAM $\mathbf{p}_{\text{ego}}$
\STATE \hspace{0.5cm}Receive updated 3D target positions $\mathbf{P}_2, \dots, \mathbf{P}_n$ from server
\STATE \hspace{0.5cm}\textbf{For each target $i = 2, \dots, n$:}
\STATE \hspace{1cm}Transform coordinates: $\mathbf{P}_i' = \mathbf{R} \mathbf{P}_i + \mathbf{t}$
\STATE \hspace{1cm}Convert $\mathbf{P}_i'$ to ego frame for visualization
\STATE \hspace{1cm}Display 3D bounding box or directional pointer in AR headset
\end{algorithmic}
\end{algorithm}

\begin{algorithm}[!ht]
\caption{AR Directional Arrow Display Logic}
\label{alg:arrow_logic}
\begin{algorithmic}
\REQUIRE Target distance $d$, relative angle $\theta$ (deg), field of view  $\mathrm{FoV}=106^\circ$
\STATE $\text{isClose} \leftarrow (d \le 3)$,\quad $\text{isInView} \leftarrow (|\theta| \le \mathrm{FoV}/2)$
\IF{$\text{isClose}$ \AND \NOT $\text{isInView}$}
    \STATE Activate arrow display (if not active)
    \STATE $\text{side} \leftarrow \text{right if } \theta>0 \text{ else left}$
    \STATE $\text{verticalPos} \leftarrow \text{up if } |\theta| < 180^\circ-\mathrm{FoV}/2 \text{ else down}$
    \STATE Place and orient arrow according to $(\text{side}, \text{verticalPos}, \theta)$
\ELSE
    \STATE Deactivate arrow display
\ENDIF
\end{algorithmic}
\end{algorithm}

\subsection{Multi-Headset Operation}

In addition to single user operation, the ARCAS framework supports multiple AR headsets sharing the same world coordinate frame. This capability is useful for collaborative scenarios such as indoor firefighting or emergency response, where several users must maintain a consistent view of nearby hazards and each other's positions.

Multi-headset calibration is performed through a simple reference point procedure. At the start of the session, all users briefly stand at the same physical location and face approximately the same direction. Each headset records its SLAM pose at this point, denoted $\mathbf{p}^{(j)}_{\text{ref}}$ for headset $j$, while one headset is chosen as the primary frame, and the remaining headsets are aligned to it.

For each additional headset $j$, a short calibration motion is performed while both headsets record synchronized 3D poses. Using the paired pose samples
$\{\mathbf{p}^{(1)}_i\}$ and $\{\mathbf{p}^{(j)}_i\}$, the system estimates a 3D rigid body transformation
$(\mathbf{R}_{j1}, \mathbf{t}_{j1})$ that maps coordinates from headset $j$ into the primary headset frame, following the same Orthogonal Procrustes formulation used for LiDAR–headset calibration:
\begin{equation}
    \min_{\mathbf{R}_{j1},\,\mathbf{t}_{j1}}
\sum_{i=1}^{n}
\left\|
\mathbf{p}^{(1)}_i -
\left(
\mathbf{R}_{j1}\,\mathbf{p}^{(j)}_i + \mathbf{t}_{j1}
\right)
\right\|^2.
\end{equation}

Once calibrated, all headsets operate in a shared world frame. Each device runs its own SLAM tracker, but its pose and detected hazards are transformed into the common frame using $(\mathbf{R}_{j1}, \mathbf{t}_{j1})$. An exclusive communication channel exchanges these transformed poses and hazard information, enabling all users to see consistent 3D bounding boxes, directional arrows, and teammate locations in real time.

\section{Experimental Protocol}

\subsection{Hardware Configuration and Physical Setup}
To evaluate the ARCAS system in a real-world setting, several experiments were conducted indoors and outdoors of the Engineering Education and Research (EER) Center at the University of Texas at Austin.
The sensing module consists of a Velodyne VLP-16, a compact 360° 3D LiDAR installed at the intersection. The sensor captures motion data in the form of 3D point clouds within a range of up to 100 meters and operates at rotation rates up to 20 Hz, providing a horizontal angular resolution as fine as 0.1° and a vertical resolution of 2.0°. These specifications enable reliable detection of small positional changes with centimeter level accuracy, supporting real-time tracking of moving targets and continuous updates to their bounding boxes.
The LiDAR was connected to a Dell G15 laptop using a Gigabit Ethernet interface, as the Velodyne VLP-16 transmits 3D point cloud data over UDP via Ethernet connection. The laptop, functioning as the system server, was equipped with an Intel 13th Gen Core i7 processor, 32 GB of RAM, and an NVIDIA GeForce RTX 4060 GPU. All point cloud processing, object detection, and Kalman Filter are performed on the CPU, and the dedicated GPU is not required for real-time operation. Based on observed computational loads, minimum requirements for real-time performance at 20 Hz are an Intel Core i7 (9th generation or equivalent) with 8 GB RAM. This setup supported real-time object detection, tracking, and wireless transmission of motion data to the AR headset over a local TCP network. 
The AR interface was deployed on a MQPro headset operating in passthrough mode. The headset uses an internal SLAM-based tracking system to determine the user spatial position and orientation. Upon receiving real-time position data from the server, the headset performs coordinate transformations from the world frame to its local reference frame and renders visual cues, such as 3D bounding boxes and directional arrows, in the user's FoV.

\begin{figure}[!t]
    \centering
    \vspace{6pt}
    \includegraphics[height=13cm, keepaspectratio]{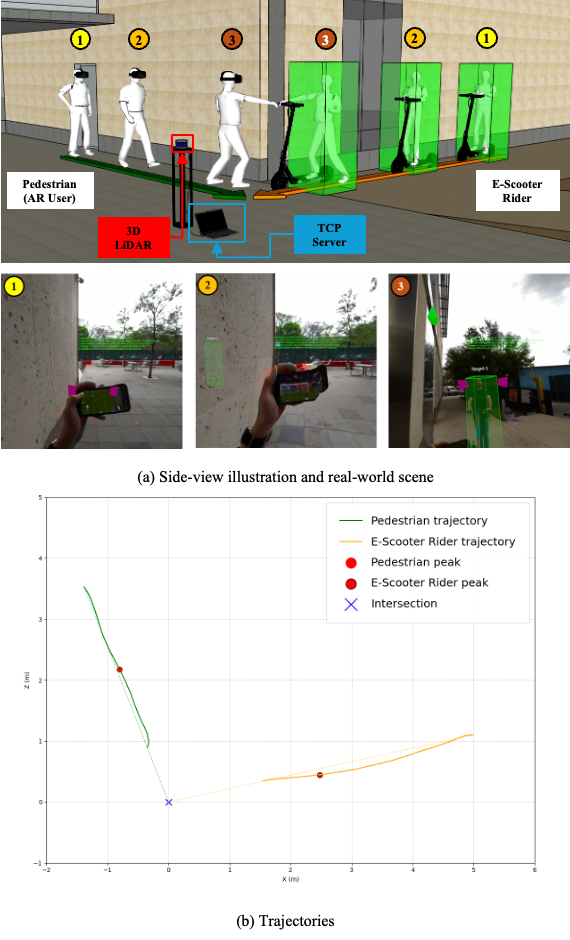}
    \caption{Overview of the pedestrian–e-scooter collision scenario.
(a) The top panel presents a side view illustration alongside the corresponding real-world scene, in which an AR-equipped pedestrian approaches an intersection while an e-scooter rider, without AR and detected solely by 3D LiDAR, emerges from behind an occluding wall.
(b) The bottom panel shows the reconstructed trajectories of the pedestrian and the e-scooter rider, including the points at which each reaches its peak velocity.}
    \label{fig:ped_scooter}
\end{figure}

\begin{figure}[!t]
    \centering
    \vspace{6pt}
    \includegraphics[height=13cm, keepaspectratio]{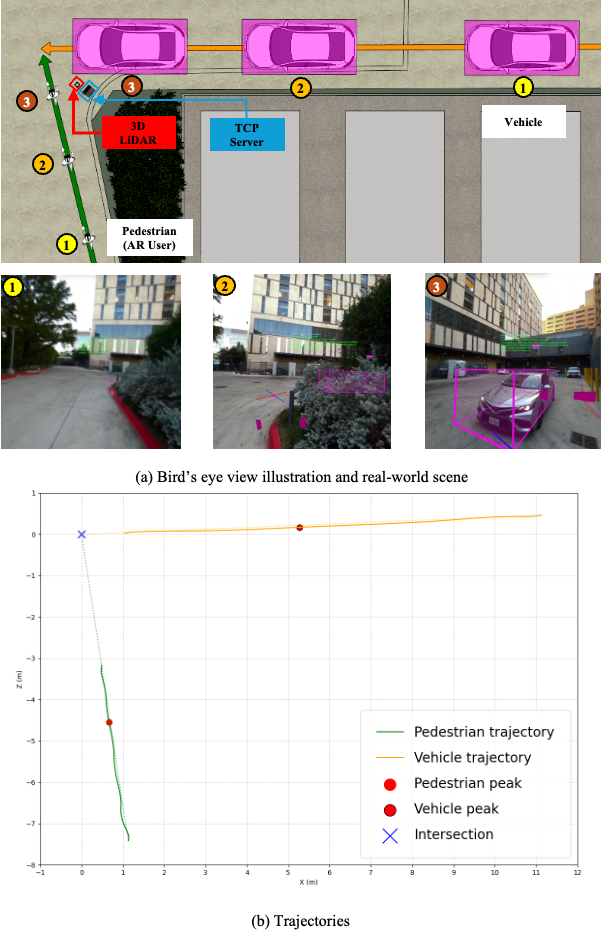}
    \caption{Overview of the pedestrian–vehicle collision scenario.
(a) The top panel shows a bird's eye illustration alongside the corresponding real-world scene, in which an AR-equipped pedestrian approaches an intersection while a vehicle emerges from an approximately 80$^\circ$ curved and visually occluded side road.
(b) The bottom panel illustrates the reconstructed trajectories of both the pedestrian and the vehicle obtained from AR headsets, including the positions at which each reaches its peak velocity.}
    \label{fig:outdoor_ped_vehicle}
\end{figure}

\subsection{Experiment Overview}


To assess the effectiveness of ARCAS in enhancing early hazard perception, two real-world outdoor experiments were conducted involving: (1) a pedestrian and an e-scooter rider, and (2) a pedestrian and a vehicle. All experiments took place in a controlled sidewalk and parking lot area outside the EER building. Each scenario was evaluated under three sensing and visualization conditions: \textit{(A) Baseline (Unaided Vision)}, \textit{(B) AR + LiDAR Detection}, and \textit{(C) AR + Multi-Headset Sharing}. For each of the two scenarios, 30 trials were performed under each of the three conditions, resulting in a total of 180 outdoor trials. The sample size of 30 trials per condition was selected to ensure stable mean estimates based on the central limit theorem. Time to collision (TTC) was used as the primary metric for evaluating early hazard perception.

Across trials, start positions and approach trajectories were held constant. Inherent differences in walking pace, reaction time, and approach speed introduced trial to trial variability. Participants were instructed to walk or drive at a comfortable pace without specific speed targets.  In the pedestrian–e-scooter scenario, pedestrian walking speed averaged 1.2 m/s (SD = 0.15) and e-scooter speed averaged 3.2 m/s (SD = 0.4). In the pedestrian–vehicle scenario, pedestrian speed averaged 1.1 m/s (SD = 0.12) and vehicle speed averaged 6.5 m/s (SD = 0.8). Conditions were counterbalanced across sessions to mitigate learning and fatigue effects.

\subsection{Experimental Scenarios}

\textbf{Scenario 1: Pedestrian--E-Scooter Interaction.}
A pedestrian and an e-scooter rider approached a shared intersection from orthogonal directions. The e-scooter emerged from a fully occluded 90$^\circ$ corner at 10--15~km/h, while the pedestrian walked toward the expected intersection point (EIP) at normal pace as shown in Figure \ref{fig:ped_scooter}. 

\smallskip

\textbf{Scenario 2: Pedestrian--Vehicle Interaction.}
This scenario involved a pedestrian crossing near a side road while a vehicle approached from a different direction. The vehicle emerged from an approximately 80$^\circ$ curved corner at 20--30~km/h, gradually entering the pedestrian’s path of motion. Figure~\ref{fig:outdoor_ped_vehicle} presents the layout of this real-world interaction. 

\subsection{Experimental Conditions}

\textbf{Condition A: Unaided Eye (Baseline).}
Neither the pedestrian nor the counterpart (e-scooter rider or vehicle driver) wore an AR device. Situational awareness depended entirely on natural human vision. Ground truth positions and velocities for TTC computation were obtained from VLP-16 LiDAR trajectories.

\smallskip

\textbf{Condition B: AR + LiDAR Detection.}
The pedestrian wore an MQPro AR headset, while the counterpart did not. The counterpart's position and velocity were estimated using the VLP-16 LiDAR, and the pedestrian received real-time AR cues (3D bounding boxes and directional arrows) corresponding to the detected hazards.

\smallskip

\textbf{Condition C: AR + Multi-Headset Sharing.}
Both the pedestrian and the counterpart wore MQPro headsets. SLAM-based multi-headset alignment provided a unified world coordinate frame, enabling direct AR to AR exchange of headset poses without relying on LiDAR. Although occasional SLAM resets occurred during fast vehicle motion, the overall tracking remained sufficiently stable for TTC analysis.

\subsection{Time To Collision Computation}

For both scenarios, TTC was used as the primary evaluation metric. At each time step, the future meeting point between the pedestrian and the counterpart was estimated using linear extrapolation of their instantaneous positions and velocities, producing the EIP.

TTC for each user is defined as:
\begin{equation}
\mathrm{TTC} = \frac{d_{\mathrm{EIP}}}{v},
\end{equation}
where $d_{\mathrm{EIP}}$ denotes the Euclidean distance from the current position to the EIP, and $v$ is the user's instantaneous speed. TTC reflects the remaining time before the user reaches the potential collision point under the current motion assumption.

\begin{table}[t]
\vspace{6pt}
\caption{Time to Collision (TTC) across different sensing and visualization configurations for two real-world near collision scenarios.}
\label{tab:ttc_vertical_full}
\centering
\begin{tabular}{l c c}
\hline
\textbf{Scenario / Configuration} & \textbf{Pedestrian [s]} & \textbf{Other Party [s]} \\
\hline

\multicolumn{3}{l}{\textbf{Pedestrian -- E-scooter}} \\
\quad Unaided Eye (Baseline)         & 2.753 & 0.694 \\
\quad AR + LiDAR Detection         & 5.512 & 2.863 \\
\quad AR + Multi-Headset Sharing   & 5.434   & 2.876  \\

\hline

\multicolumn{3}{l}{\textbf{Pedestrian -- Vehicle}} \\
\quad Unaided Eye (Baseline)         & 5.081 & 0.855 \\
\quad AR + LiDAR Detection         & 6.421 & 1.984 \\
\quad AR + Multi-Headset Sharing   & 6.285   & 1.889   \\

\hline
\end{tabular}
\end{table}

\subsection{Summary of Results}

Table~\ref{tab:ttc_vertical_full} presents the average TTC values computed across 30 repeated trials for each configuration, quantitatively demonstrating that ARCAS substantially improves early hazard perception across both scenarios. Several key findings emerge. Across all conditions, TTC values for the pedestrian increased substantially under AR guidance. In the pedestrian–e-scooter scenario, TTC increased from 2.75 s in the Unaided Eye baseline to 5.51 s with AR + LiDAR and 5.43 s with multi-headset sharing, indicating an approximately twofold improvement in the pedestrian’s available reaction time. The e-scooter rider also exhibited substantial gains, with TTC increasing from 0.69 s to approximately 2.86 s under AR conditions, more than a fourfold improvement. In the pedestrian–vehicle scenario, AR + LiDAR Detection produced the highest pedestrian TTC (6.42 s), while multi-headset Sharing delivered a similar improvement (6.29 s) despite occasional SLAM resets. Overall, these results confirm that ARCAS, whether powered by external LiDAR sensing or AR to AR SLAM sharing, provides significantly earlier hazard awareness and larger safety margins for VRUs in occluded and fast approaching traffic conditions.

\section{Conclusion}

This paper introduced ARCAS, a real-time \textbf{A}ugmented \textbf{R}eality \textbf{C}ollision \textbf{A}voidance \textbf{S}ystem designed to enhance VRUs’ situational awareness in dynamic traffic environments. By integrating 360° 3D LiDAR sensing, SLAM-based headset tracking, and automatic 3D calibration, the system delivers world locked visual cues, such as 3D bounding boxes and directional arrows, directly in the user’s passthrough view. ARCAS further supports multi-headset alignment, enabling multiple VRUs to share a unified world frame for coordinated awareness. From a broader perspective, ARCAS represents a step toward an intersection-level digital twin; however, a full digital twin implementation would require additional components such as a persistent world model, and scenario replay capabilities, which remain directions for future development.
The system was evaluated through two real-world near collision scenarios involving pedestrian interactions with an e-scooter rider and a vehicle. Across 180 trials, AR visualization consistently increased TTC for both the pedestrian and the counterpart. In the pedestrian–e-scooter scenario, the pedestrian’s TTC nearly doubled and the rider’s TTC increased more than fourfold under AR guidance. In the pedestrian–vehicle scenario, LiDAR-based detection yielded the greatest improvements, while multi-headset sharing achieved similar gains despite occasional SLAM resets. These results demonstrate that AR-based visual cues, whether from external sensing or AR to AR pose sharing, provide substantially earlier hazard awareness and larger reaction windows in occluded or fast approaching conditions.

Overall, ARCAS establishes a practical foundation for AR-based safety systems that deliver real-time, personalized hazard awareness for VRUs. Future work will explore intent prediction, adaptive cue generation, and context aware filtering, with expanded evaluation using additional surrogate safety measures such as post encroachment time and deceleration rate to avoid collision. Future studies will also address human factors including user acceptance, cognitive workload, and trust in wearable AR for traffic safety. Moreover, the designed current experiments were conducted in a controlled campus environment, and future work should evaluate ARCAS in more complex and dense urban traffic settings to advance intelligent, human-centered safety solutions for next generation mobility systems.

	\section*{ACKNOWLEDGMENTS}
This research is partially supported by a grant from Honda Development \& Manufacturing of America, LLC (HDMA).
	
	\bibliographystyle{IEEEtran}
	\bibliography{root} 
	
\end{document}